\documentclass[10pt,journal,compsoc]{IEEEtran}
\usepackage[T1]{fontenc}
\usepackage{amsmath,amssymb}
\usepackage{graphicx}
\usepackage{url}
\usepackage{booktabs}
\usepackage{array}
\usepackage{ragged2e}
\usepackage{textcomp}
\begin{document}
\title{What a Policy Gate Can and Cannot Know: Measured Boundaries of Cross-Platform Command Adjudication}
\author{Qishuai Jing\thanks{The author is an independent researcher in Yangzhou, Jiangsu 225001, China (e-mail: jqs113@foxmail.com). ORCID: 0009-0007-3951-1763. Corresponding author. This work has been submitted to the IEEE for possible publication. Copyright may be transferred without notice, after which this version may no longer be accessible.}}
\maketitle

\begin{abstract}
Gateways that adjudicate an agent's actions before they execute are only as good as their understanding
of the action. We study a policy gate that never parses shell syntax: it consumes a typed, realised action
(verb, operands, resolved zones, program-object identity) and decides ALLOW, ASK or DENY. Working on a Linux
twin of the Windows benchmark of our previous study~\cite{prev}, we ask how faithfully it adjudicates, what
survives translation, and whether deciding stays affordable as the system is used. A frozen 61-case table
scores 61/61 in two rounds with no false allow; a 50-operator mutation campaign kills 46 of 50 mutants
(92.0\%), with all four surviving mutants classified. An 82-row audit yields 43 re-expressions, 21 carrier
differences, 16 study-specific inapplicable rows and two unresolved cases; of 25 rows labelled ``no
counterpart,'' four remain unmapped here.
Consulted by the executor, 25 escapes become zero with no benign payload blocked. In an exploratory
one-gateway snapshot, three unauthenticated endpoint labels produce case-level non-refusal majorities of
81.8--98.0\%, but immediate-execution majorities of 4.0--52.5\%. Adjudication reads no accumulating state;
credential verification does, scanning its whole ledger. We report that cost and the fix we would make.
\end{abstract}


\section{Introduction}
\label{sec:intro}

A gate between a proposed action and its privileged execution decides nothing about the \emph{intent}; it decides about the \emph{action}. That makes its fidelity measurable in a way intent classifiers are not: for a given action and policy state the correct decision is a matter of record, and every disagreement is a defect with an address. It also makes a different class of question askable: not ``how often is the verdict right'' but ``what does this component know, and where does that
knowledge stop''. This paper is about the second question.

This paper is the second half of a two-part study. Part one~\cite{prev} introduced the governed execution
architecture and evaluated it on a Windows benchmark of 313 cases; its subject is the architecture and the
Windows results, and \emph{we do not re-report them here}. What it left open is the question this paper answers: when the same gate is moved to Linux, does it still work, what transfers, and --- more useful to future builders --- \textbf{what are the boundaries it cannot cross}. The shared platform (the corpus, the decoy tree, the fingerprint discipline) is therefore described only as far as our arguments require, and each borrowed element is cited rather than restated. We treat the gate as an instrument, and report its own failure modes as results.

Three questions organise the paper.

\textbf{RQ1 (fidelity).} How faithfully does a gate that consumes structured actions adjudicate, and how
much of that fidelity is a property of the gate rather than of the cases it was given? A frozen table can be passed by a rule set tuned to it, so the interesting measurements ask what the corpus would \emph{not} notice: mechanism tables whose claims are made in two carriers each, and a mutation campaign that reports which defects survive rather than how many were caught.

\textbf{RQ2 (portability).} Which governance rules survive translation from one platform to another, and
which cannot be mapped under a defined target host and toolset? The point is not the count but the reason:
answering requires judging rules by the capability they express, not the name they carry, and reporting
lossy or absent mappings without treating absence from one image as proof of platform-wide nonexistence.

\textbf{RQ3 (execution).} Does the gate change what actually happens on the machine, and what does it cost
in false interception? A decision procedure that nothing consults is a document. We therefore measure both
directions at once: escapes with the gate in the path versus out of it, and benign payloads that must not
be blocked --- which turns out to constrain the enforcement policy more than the escape count does.

The contributions are these.

\begin{itemize}
\item \textbf{A fidelity measurement of the adjudicator itself} (Section~\ref{sec:rq1}): a frozen 61-case
table at 61/61 in both rounds, six mechanism tables (78 rows) that make every claim in two carriers, and a
50-operator mutation campaign with a 92.0\% mutation score (46/50 mutants killed) and all four surviving
mutants classified --- three as corpus blind spots named individually, one as a property of the method.
\item \textbf{A cross-platform study judged by capability rather than by name} (Section~\ref{sec:rq2}). Its
82-row audit pool combines 57 manual-review candidates with 25 rows originally labelled ``no Linux
counterpart.'' Across the pool, 43 are re-expressed, 21 differ by carrier, 16 are not instantiated under our
mapping, and two remain unresolved. Within the original 25, four rows remain unmapped. Separately, one row elsewhere in the pool is
translated faithfully and yet inverts its meaning rather than disappearing. A
98-row comparison sheet, authored for this paper, records 53 native and 45 carrier-differing rows.
\item \textbf{An execution-side experiment with both directions measured} (Section~\ref{sec:rq3}): 25
escapes become zero over 65 payloads, with zero false interception, and with the interception policy
justified by measurement rather than by default.
\item \textbf{An exploratory same-case endpoint snapshot} (Section~\ref{sec:llm}): on the subset whose
denial depends on nothing platform-specific, three unauthenticated labels have non-refusal majorities of
81.8\%, 94.9\% and 98.0\%, whereas immediate-execution majorities are 24.2\%, 4.0\% and 52.5\%.
No label refuses a gate-ALLOW case; answers also change when the case is rewritten for another platform.
\item \textbf{Three methodological findings that generalise beyond this instrument} (Sections~\ref{sec:method}
and~\ref{sec:threats}): a reading that depends on the live filesystem is a function of the range state and
must be reported with it; inherited payloads must be graded for trustworthiness before any false-allow
count is interpreted; and a conservative rule may be doing a second job --- the rule that escalates every
action without a path target to review was also catching target-extraction failures, which is invisible
until the rule is relaxed and an inherited case turns from review into permission.
\item \textbf{Negative and falsified results, reported as such.} Five predictions in this study were
falsified by data, and the boundaries in Section~\ref{sec:limitations} include what the layer cannot do at
all: propagate state across sequential statements, resolve a symbolic name without a path, or see effects
outside the filesystem.
\end{itemize}

\section{Related Work}
\label{sec:related}

\textbf{Policing the input.} A large body of work hardens the path \emph{into} the model: taxonomies and
benchmarks for prompt-injection attacks and defences~\cite{liu24}, indirect-injection benchmarks for
tool-using agents~\cite{zhan24}, dynamic evaluation environments~\cite{debenedetti24,zhang25}, and designs
that restructure the interface between data and instructions so the two cannot be confused by
construction~\cite{chen25struq,debenedetti26,bk25}. This work is complementary and, in one sense, orthogonal to ours: the gate never sees model input, and every case in our corpora is adjudicated on the action proposed. Where the two meet is a question we do not answer --- how an action came to be proposed --- and we scope it out explicitly.

\textbf{Interposing a component.} A second line places an adjudication or isolation point between the
agent and the system: execution isolation for agentic systems~\cite{wu25}, security architectures for
applications that integrate large language models (LLMs)~\cite{li26} and for agent governance~\cite{syros26}, customisable runtime
enforcement~\cite{wang26}, programmable privilege control~\cite{shi25}, and capability-tracked
agents~\cite{odersky26}. Industry control frameworks now assume such a component exists: access-control
decisions must be enforced by application logic or a policy engine rather than by the
model~\cite{owasp26}, agent identities must be governed~\cite{csa26}, and separation of duties together
with audit protection is standard control vocabulary~\cite{nist25} under a general risk-management
umbrella~\cite{nist23}. What this line establishes is that an external decision point is buildable; what
it does not report is the \emph{fidelity of the adjudicator itself} --- how often it is right for the
right reason, and where the boundary of its knowledge lies once the platform changes beneath it. That is
what this paper measures.

\textbf{Testing the policy, not the program.} Mutating an instrument and scoring a corpus with it is an
old idea with a policy-specific literature: Martin and Xie define a fault model and mutation operators
for access-control policies~\cite{martinxie07}, Mouelhi \emph{et al.} generalise the operators across
rule-based policy formalisms~\cite{mouelhi08}, and Jia and Harman survey the
field~\cite{jia11}. Our campaign follows that logic --- the object under test is the decision procedure,
not the workload --- with one addition that our own results forced: comparing verdicts alone cannot kill
a change that alters only the \emph{reason} text, so we add a reason layer whose signature is the verdict
together with the reason that decided it. Six such changes survived a verdict-only campaign in an
earlier round.

\textbf{Capabilities, names, and why portability fails.} The classical statement is that a subject's authority should be minimised and every access mediated~\cite{saltzer75}, that ambient authority produces the confused deputy~\cite{hardy88}, and that unforgeable references are the structural remedy~\cite{capsicum10}; on Linux, self-restricting sandboxes carry the same idea~\cite{landlock}. RQ2 is, in these terms, an empirical measurement of how much of a rule set phrased
in names and paths survives when its carrier changes. Mapping status is bounded by the target host and
toolset: most rows are re-expressed, four of the 25 originally labelled ``no counterpart'' remain unmapped
in our corpus, and one translation changes meaning. None of these corpus labels proves platform-wide absence.

The architecture studied here, together with its Windows evaluation, is the subject of our previous
study~\cite{prev}. The present paper is its cross-platform and instrument-fidelity counterpart, and
neither repeats the other's results.

\section{System: a gate that consumes realised actions}
\label{sec:system}

\begin{figure*}[t]
\centering
\includegraphics[width=0.94\textwidth]{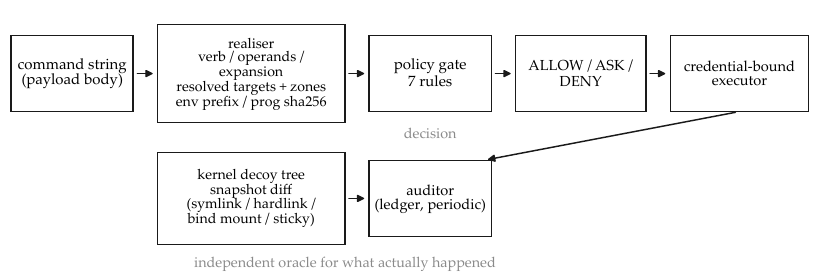}
\caption{The adjudication chain. The gate consumes a realised action --- verb, operands, expansion
count, resolved targets with their zones, environment prefix and program-object digest --- and never
shell text. The auditor and the decoy-tree oracle sit outside the executor.}
\label{fig:chain}
\end{figure*}

The adjudication chain is seven files, all of which are hashed into a single \emph{instrument fingerprint}
reported with every number in this paper. Two components do the work that matters here (Fig.~\ref{fig:chain}); a third supplies
the authority, and a fourth exists because not every action starts life as a command string.

\subsection{The realiser}
The realiser turns a command string into a \emph{realised action}: a status, a verb, the operands, an
expansion count, path targets with their resolved zones, an environment prefix, and --- when the program is
given by path --- the \texttt{sha256} of its bytes. It is deliberately conservative. Constructs it cannot
statically enumerate (pipelines, subshells, command substitution, brace groups) yield a
\textsc{Multiproc\_Undecidable} status rather than a guess, and sequential actions separated by a
semicolon or a newline are split, adjudicated one by one, and then aggregated: any denial dominates, and
more than one distinct target object escalates to review. Two details of the design were themselves
measured rather than assumed. First, the expansion count records \emph{how many targets an expansion added},
not whether a pattern was used --- a glob that matches once and a variable substitution are both invisible,
because after the shell has expanded the word the evidence is gone. Second, plurality and expansion are
complementary rather than redundant: the only shape in which expansion alone decides is a \texttt{cp} or
\texttt{mv} whose glob expands to exactly two sources, because a fixed-arity verb pair legitimately takes
two paths and the plurality signal cannot see the difference.

\subsection{The gate}
The gate applies seven rules to each realised action: fail-closed on unrealisable input, including the
gate's own internal exceptions; sensitive verbs, which are demoted by effect rather than by name;
program identity --- by name, by position, and by object content and attributes; zone judgement on
resolved targets; protection of the workspace root itself; expansion and plurality; and finally intent
binding, which requires the declared object set to cover the set of objects actually acted upon. Rules are
not independent and the dependencies are deliberate: the zone rule resolves symbolic links before judging,
so a path that looks in-workspace and an object that is not in-workspace are different answers by design.

\subsection{Zones, credentials, and the intent channel}
The range defines five zones: a writable working area whose only defence is the gate; a zone writable by
the test identity but treated as protected, which is how the gate's efficacy is separated from the
kernel's; a root-only zone, which tests the DAC (discretionary access control) layer underneath; an outside-writable zone; and everything
else, which is foreign. Authority is not implied by a decision: an approval issues a one-shot credential
bound to the verb, the object identities (device and inode), the paths, the full argument vector, the
argument keys and the content digest, and it is consumed exactly once, so a replayed or mutated request
fails even if the action would otherwise be permitted. Actions may also arrive as \emph{structured
intents} rather than command strings --- a typed operation, absolute paths, typed arguments --- in which
case adjudication never parses shell syntax at all; the two channels share the gate and the credential
layer, and Section~\ref{sec:rq1} reports what happened when their producers disagreed about a field.

\subsection{Three decisions the rest of the paper measures}
\emph{Zones, not names:} the gate asks where a target resolves to, not what it is called, which is why the
cross-platform question is answerable at all. \emph{Quote evidence:} a word containing whitespace or shell
metacharacters is treated as a string rather than as a path target, because the evidence that it was quoted
is still present at adjudication time. \emph{Explicit admissions:} when the gate cannot read a construct it
says so in the reason text, because a reason the caller cannot act on is a defect rather than a detail ---
Section~\ref{sec:rq1} reports one such case being fixed, and Section~\ref{sec:limitations} one such case
being paid for.

\section{Measurement methodology}
\label{sec:method}

The object of study is an instrument, so the way it is measured is part of the result. The artifact list, the corpora and the recovery procedure are in Section~\ref{app:artifacts}. Five disciplines
were adopted early and each of them later changed an outcome; we describe them with the case that did.

\subsection{M1: measure, then predict, then change the code}
Every change to the chain starts as a written prediction containing assertions that can be true or false,
and the prediction is archived whether it holds or not. Three of them changed what we did rather than
confirming it: a claim that a change fixed a decision became a claim that it repaired only the attribution,
because measuring first showed two reasons contributing to the same decision; a relaxation believed to be
the mechanism by which benign script payloads could pass turned out to change no decision at all across the
65 payloads and the frozen table; and a self-test expectation written from a guessed classification failed
46 of 49 immediately and named the entry that was wrong. The prediction archive is in the supplement,
including the five that were falsified.

\subsection{M2: two oracles at three layers}
\begin{figure*}[t]
\centering
\includegraphics[width=0.94\textwidth]{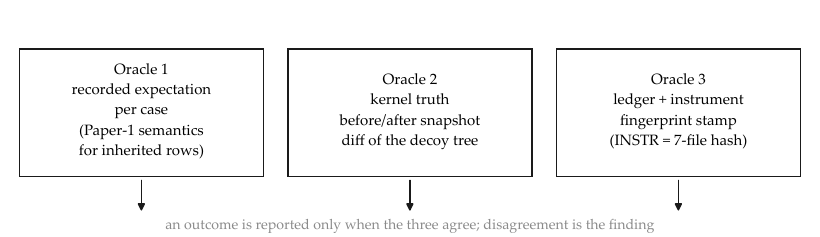}
\caption{Three oracles. A recorded expectation, a kernel-level before/after snapshot diff of the decoy
tree, and an instrument-fingerprint-stamped ledger. An outcome is reported only when they agree;
disagreement is treated as a finding about the payload, the range, or the gate, in that order.}
\label{fig:oracles}
\end{figure*}

Kernel ground truth comes from a structured decoy tree --- symlinks, hard links, bind mounts, sticky
directories --- snapshotted before and after execution and diffed (Fig.~\ref{fig:oracles}); the expected decision recorded per case
and the fingerprint-stamped ledger are the other two layers. A report that only counts decisions cannot
see an escape: in the enforcement experiment (Section~\ref{sec:rq3}) the 25 escaped payloads are escapes
precisely because their side effects were observed, not because the gate said something. The oracle has a
residual of 2.6\% (3 of 117 runs), and it is a property of the method rather than a defect of the gate:
an object created \emph{during} execution cannot appear in a pre-execution snapshot, so an action that
both creates and aliases a new object can alter the tree without the diff attributing it.

\subsection{M3: a denominator ledger}
Not every inherited case may serve as a denominator. The 313 source cases are a comparability anchor: they
keep this study's numbers commensurable with the earlier one, but only the rows with a Linux spelling are
adjudicable, and of those, a difference from a recorded expectation is not automatically an error. Reporting
a rate over the whole set would be arithmetic without meaning. This discipline paid for itself
immediately, because it is what caught a vacuous round: the declared-object column of the mapping table
had been derived from the commands themselves by regular expression, with every non-empty value appearing
literally inside its own command, so the intent-binding rule could never fire and the with-declarations
round was measuring something identical to the without-declarations round. A number that looks like a
result can be a tautology, and only the constitution of the denominator exposes it.

\subsection{M4: two-layer mutation, with reason probes}
Mutating the gate and comparing verdicts is not sufficient, because a change that alters only the reason
text leaves every verdict intact. Comparing verdicts only, six such changes survived; adding a reason layer
killed five of them, and two positive controls confirm that the layer carries load rather than decoration
(Section~\ref{sec:rq1}). The operator table also carries a source-hash guard so that an operator whose
anchor has drifted fails loudly instead of measuring something else --- an anchor failure is a number that
looks plausible, not an error message.

\subsection{M5: declared range state}
Readings that depend on the live filesystem are reported together with the state of the target range they
were taken in, because they are functions of it. Deleting a single decoy file moves the mutation score
from 92.0\% to 94.0\% and changes one operator's fate, since the signal it removes is produced by glob
expansion against a real directory. Two readings taken in different range states are two measurements of
two things, and treating them as one --- which is what an inherited number invites --- produced the single
most misleading comparison of this study (Section~\ref{sec:threats}).

\subsection{Failures are reported}
Five predictions in this study were falsified by data, and each is reported where it happened rather than
in a list at the end: one turned a change into a different change, one was a mis-specified self-test
expectation, one inherited a stale baseline from a previous hand-over, and one was a cross-tabulation of
ours caught by recomputing it cell by cell, and one is the cost of a growing ledger, read by us as
per-byte work where the measurement says per-row (Section~D.1 of the supplement). A methodology that reports only its successes cannot be
evaluated.

\section{RQ1: fidelity of adjudication}
\label{sec:rq1}

\begin{figure*}[t]
\centering
\includegraphics[width=0.94\textwidth]{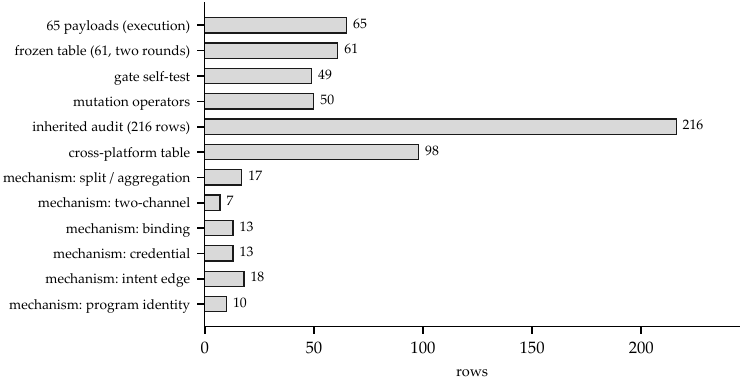}
\caption{Corpora used by the two research questions. Corpus sizes are not interchangeable: the
mechanism tables and the frozen table may serve as denominators, whereas the inherited rows and the
cross-platform sheet serve coverage auditing only.}
\label{fig:corpora}
\end{figure*}

\textbf{Frozen table.} 61 cases, two rounds (with and without declared intents; corpus sizes in Fig.~\ref{fig:corpora}):
\textbf{61/61 and 61/61}, false-allow 0.0\%, false-deny 0.0\%. The gate's own self-test is 49/49 and
the credential layer's is 12/12.

\textbf{Mechanism tables.} Six purpose-built tables, 78 rows in total, test one mechanism each
(Table~\ref{tab:mech}). Three properties distinguish them from a pass-rate corpus. First, every claim is
made in at least two carriers: the same effect reached by different spellings must land in the same
decision, and the same spelling must reach different decisions when the \emph{effect} differs --- so a
rule that fires on syntax rather than semantics fails visibly. Second, cross-row assertions are made
where single-row comparisons cannot see the mechanism: that four spellings of one declaration resolve to
the same object, that an exemption is directional, that a copy's destination is not exempt while a move's
is, that the object count is unchanged across a split. Third, every suite that feeds deliberately broken
input also feeds a positive control, because a suite that only feeds bad input can be satisfied by a layer
that rejects everything.

The tables earned their keep in what they caught, and the list is worth stating plainly: a table
separator ate a pipe character and silently dropped an entire row (which also meant piped constructs had
no case anywhere --- they lived only in in-code self-tests); two different quoting conventions coexisted
inside one system, so a semicolon inside quotes fell into a gap that was neither split nor judged; and
two producers of the same action type disagreed about the \emph{role} of a field, which made the same
request adjudicate oppositely depending on the channel it arrived through. The 313-case inherited corpus
caught none of these three, because none of them is a decision about a command --- each is a decision
about the instrument.

\begin{table*}[t]
\caption{The six mechanism tables (78 rows). ``Axis pinned'' is the design question the block exists to
answer; ``caught'' lists the defect the block actually found.}
\label{tab:mech}
\centering
\small
\begin{tabular}{>{\RaggedRight\arraybackslash}p{1.8cm}c>{\RaggedRight\arraybackslash}p{4.6cm}>{\RaggedRight\arraybackslash}p{6.2cm}}
\toprule
block & rows & axis pinned & caught \\
\midrule
mech & 17 & sequential split and aggregation & separator ate a pipe: one row silently dropped \\
chan & 7 & string channel vs structured intent & producers disagreed on \texttt{role} (same request, opposite verdicts) \\
bind & 13 & binding on object identity & relative-declaration normalisation \\
cred & 13 & one-shot credential & two layers bind different things \\
edge & 18 & intent-layer byte/JSON/vocabulary/path edges & a layer that refuses everything passes a one-sided suite \\
prog & 10 & program identity by position and by content & a renamed copy is recognised by content, not by name \\
\bottomrule
\end{tabular}
\end{table*}

\textbf{Mutation campaign.} We mutate the decision procedure and score the corpus with it: 50 semantic
operators over four corpora --- the gate's own self-test (49), the frozen 61-case table plus its separate
13-row supplement evaluated without declarations (74) and with them (74), and three fault-injection cases
--- plus 17 reason-layer probes (Table~\ref{tab:mut}).
The object under test is the instrument, not the workload, which is what makes the campaign say something
the case tables cannot: it reports what the corpus \emph{would not notice}.

\begin{table}[t]
\caption{Mutation campaign, evaluated at fingerprint \texttt{285ec6bd7ecc}. ``Spec kills'' are mutants
that turn a case the unmutated gate judged correctly into a wrong judgement.}
\label{tab:mut}
\centering
\footnotesize
\begin{tabular}{lc}
\toprule
quantity & value \\
\midrule
operators (effective mutants) & 50 \\
killed by specification & 41 \\
\quad of which unsafe (deny$\rightarrow$allow) & 24 \\
killed by the reason layer alone & 5 \\
killed in either layer & 46 \\
mutation score & 92.0\% \\
dangerous-operator score & 48.0\% \\
survivors & 4 \\
anchor failures (source-hash guard) & 0 \\
unmutated gate correct on campaign rows & 198/200 \\
positive controls (fault reaches the part under test) & 3/3 \\
\bottomrule
\end{tabular}
\end{table}

\textbf{Why a reason layer is necessary.} An earlier round compared verdicts only, and six changes that
alter the reason text while leaving every verdict intact survived it. Adding reason probes --- each
signature being the verdict, the reason that decided it, and the per-class reason counts --- kills five
of them, and the two positive controls are the evidence that the layer carries load rather than adding
decoration: both are changes that alter only the wording of a reason, both leave the verdict layer at
difference zero, and each is caught by exactly one probe. The probes are themselves self-checked against
pre-written expectations (17/17), so a probe that silently stops firing is reported instead of being
counted as a pass.

\textbf{The four survivors, classified.} Three are corpus blind spots, and naming them is the point of
running the campaign: the privileged-tool list omits \texttt{su}; zone matching is invariant between
longest-prefix and shortest-prefix on these rows, so a mutant that swaps them is not distinguished; and a
declaration resolved by \texttt{abspath} rather than \texttt{realpath} is not distinguished either, because
no row declares a path that traverses a symbolic link. The fourth survivor is different in kind and we
report it as a property of the \emph{method}: the operator drops the expansion signal in favour of the
plurality signal, and whether it is killed depends on the state of the target range, because the signal
it removes is produced by glob expansion against the live filesystem. Deleting a single decoy file
converts its status (Section~\ref{sec:threats}).

\textbf{A guard against silent staleness.} The operator table records the text hash of the source it was
written against and refuses to run when the file does not match. This is not ceremony: an operator whose
anchor has drifted measures something other than what it claims, and the failure mode is a plausible
score rather than an error message.

\textbf{Auditing inherited payloads.} The 313 inherited cases are command strings, and their recorded
expectations encode the semantics of the platform they came from. We therefore treat the audit's first
product as \emph{payload trustworthiness} rather than gate quality: of the 216 rows that have a Linux
spelling, all 216 adjudicated without a crash (191 single actions, 15 statically undecidable, 10 split
into sequences), with verdicts DENY 132, ALLOW 43, ASK 41. Agreement with the recorded expectation is
133; the 83 differences are attributed rather than scored (58 other, 19 out-of-zone targets, 3
workspace-root, 2 plurality, 1 fail-closed).

The one red line is the false-allow set, 30 rows. Grading it by cause is what makes it interpretable:
11 are carrier misalignment (the attack lives in a document or a script for which the derivation had
only a reference command), 10 are cross-platform mechanism differences (the source platform protects by
name and allow-list, this one by zone), and 9 are mapping errors in the derivation itself, where a
relative escape or a placeholder was anchored into the workspace and thereby turned a protected file
into an in-workspace file. Only the middle class can be a blind spot of the gate, and the three are kept
apart. (An internal prediction written before the audit expected zero false allows; the measurement contradicted that
expectation, the value is 30, and we cite the measurement.)

\textbf{Multi-line smuggling, on payloads we did not author.} Six inherited rows carry a second action on
a new line --- the most realistic smuggling shape in the corpus, and the first time sequential splitting
was exercised by payloads outside our own case tables. Segments per row are 2/2/2/2/4/2 and four of the
six are denied. The two that are not are explained: in one, the derivation anchored the guard path into
the workspace, so the second action never leaves the zone; in the other, the smuggled action lands in the
install tier and escalates to review instead of a denial. Both explanations concern the payload or the
policy tier, not the split.

\textbf{A repaired fail-open.} One prediction failure exposed a real hole: \texttt{bash -c}-based
expansion swallowed the exit status of a failed parse, so a syntactically broken payload fell back to an
empty argument vector, leaving the gate with no target to judge --- and therefore with nothing to deny.
The fix exits non-zero on parse failure, which converts silent allowance into a refusal. The
corresponding false-allow count in the inherited audit fell from 42 to 30 over the same series of
changes.

\section{RQ2: what crosses platforms}
\label{sec:rq2}

The audit pool contains 82 source-rule rows: 57 requiring manual confirmation (31 whose paths did not land
in the range and 26 whose verbs missed the translation table), plus 25 originally labelled ``no counterpart
on Linux.'' We re-judged them by \textbf{capability semantics} within the chosen Ubuntu image and toolset.
Across the pool, 43 rows are re-expressed as Linux shapes, 21 are carrier differences or already covered,
16 are not instantiated or applicable under this study's mapping, and two credential-layer rows whose paths
never landed in the range remain unresolved ($43+21+16+2=82$). These are study-specific mapping statuses,
not claims that a platform lacks the capability.

\begin{figure*}[t]
\centering
\includegraphics[width=0.94\textwidth]{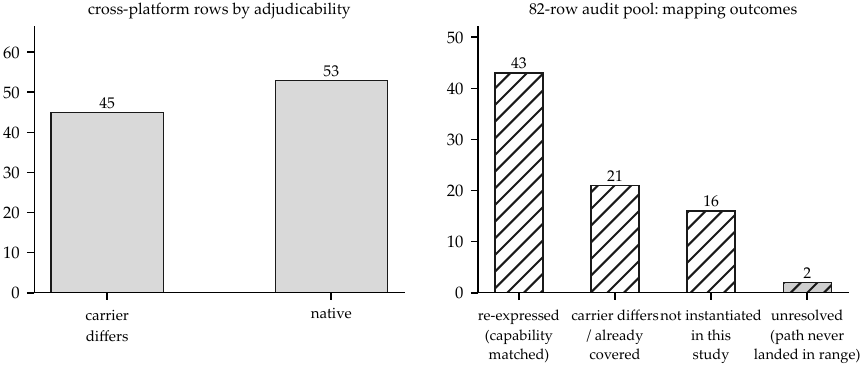}
\caption{Left: the 98-row comparison sheet by adjudicability. Right: the separate 82-row audit pool
(57 manual-review candidates plus 25 rows originally labelled ``no counterpart''), re-judged within this
study's Ubuntu image and toolset. Categories are mapping outcomes, not platform-wide capability claims.}
\label{fig:xplat}
\end{figure*}

Within the 25 rows originally labelled ``no counterpart,'' only four rows remain unmapped in this corpus:
\texttt{reg query}, \texttt{reg add}, \texttt{reg delete}, and \texttt{wmic}. They refer to Windows registry
or WMI (Windows Management Instrumentation) object models for which this study defined no Ubuntu-native mapping. That bounded result does not
exclude analogous administrative capabilities or mappings under another host or toolset. Meanwhile
\texttt{Get-Disk} becomes \texttt{lsblk}, \texttt{taskkill} becomes \texttt{pkill}, \texttt{schtasks}
becomes \texttt{crontab}, and a path composed from character codes or reversed with \texttt{rev} is
re-expressed readily: the capability transfers while the carrier changes. Table~\ref{tab:xplat} summarises the sheet by axis, and Fig.~\ref{fig:xplat} shows both pools by outcome.

\begin{table}[t]
\caption{The cross-platform comparison sheet: 98 rows, all authored for this paper. ``Native'' rows are
judged as written; ``carrier differs'' rows exercise the same capability through a different mechanism.}
\label{tab:xplat}
\centering
\footnotesize
\begin{tabular}{lcc}
\toprule
axis & native & carrier differs \\
\midrule
execution (EXC) & 6 & 8 \\
resource (RSV) & 22 & 8 \\
capability (CAP) & 6 & 0 \\
permission (PRM) & 11 & 21 \\
delete (DEL) & 4 & 0 \\
write (WRT) & 4 & 8 \\
\midrule
total & 53 & 45 \\
\bottomrule
\end{tabular}
\end{table}

\textbf{One rule inverts rather than disappears.} Windows environment-variable names are
case-insensitive; Linux ones are not. A rule that protects a path by expanding \texttt{\%windir\%}
therefore does not merely lose its carrier when translated --- it protects something else, because the
lower-case spelling that the source rule treats as the same variable names a different (or empty) one
here. We record the inversion as a positive result for RQ2 rather than filing it under ``not
applicable'': it is the cleanest example in the corpus of a translation that is syntactically faithful
and semantically opposite.

\textbf{Two projections are lossy and are stated as such.} Rules that protect by \emph{name} --- a
version-control directory, a project-specific directory --- have no zone-level equivalent in this gate's
model, so ten inherited rows differ from their expectations for that reason alone. Separately, UNC (Universal Naming Convention) paths
cannot be made faithful without inventing a mount policy, so the translation of those rows is
acknowledged as lossy rather than presented as equivalent.

\section{RQ3: does it change what happens on the machine?}
\label{sec:rq3}

The gate is a decision procedure; the question is whether anything consults it. We therefore added an
enforcement hook to the execution driver --- behind an explicit switch, off by default so that every
historical measurement stays reproducible --- and compared passes over the same 65 payloads under four
interception policies, each pass taken on a freshly restored target range (the per-case record is
Table~S4 of the supplement). Table~\ref{tab:p3} and Fig.~\ref{fig:p3} are the paper's most
direct statement: with the switch on, \textbf{25 escapes become zero and no benign payload is blocked}.

\begin{table*}[t]
\caption{Execution-side enforcement over 65 payloads under four interception policies. Each pass starts
from a restored range image, a regenerated decoy tree and the two restored program fixtures; the
control and \texttt{DENY}-only arms were re-measured at fingerprint \texttt{6c083e60ceeb} and reproduce
their earlier counterparts at \texttt{285ec6bd7ecc} byte for byte. \texttt{DENY} and \texttt{DENY+ASK}
name what the hook intercepts;
\emph{stripped} removes the payload generator's template shell before the gate sees the text. See Section~B.1d for the \texttt{AppArmor} column.}
\label{tab:p3}
\centering
\footnotesize
\begin{tabular}{>{\RaggedRight\arraybackslash}p{4.1cm}ccccc}
\toprule
Outcome & Control & \texttt{DENY} & \texttt{DENY+ASK} & \texttt{DENY+ASK}, stripped & \texttt{AppArmor} \\
\midrule
ESCAPE (DENY expected, effect observed) & 25 & 0 & 0 & 0 & 8 \\
UNDER\_\allowbreak{}BLOCK & 32 & 57 & 57 & 57 & 49 \\
Executed as allowed (ALLOW group) & 8 & 8 & 0 & 8 & 8 \\
False interception of ALLOW group & --- & 0 & 8 & 0 & 0 \\
\bottomrule
\end{tabular}
\end{table*}

\begin{figure}[t]
\centering
\includegraphics[width=\columnwidth]{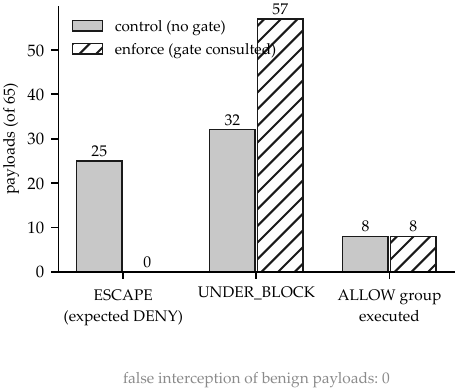}
\caption{Enforcement over the same 65 payloads, both passes on a restored range. In either pass the gate blocks the benign group zero times, unlike the strict policy of Table~\ref{tab:p3}.}
\label{fig:p3}
\end{figure}

\textbf{The gate must be fed the payload body.} Feeding it the invocation --- \texttt{bash payload.sh} ---
would show it a script path and nothing else, and any interception obtained that way would be a
\emph{fake} interception. The hook reads the payload's text and adjudicates it with the working
directory set to the same path the executor uses, because relative paths resolve differently otherwise.

\textbf{Only DENY is intercepted, and this too is measured rather than assumed.} In this corpus the gate
returns ASK for the shebang line and for comment lines at the top of a payload, so all eight benign
payloads carry ASK, and intercepting ASK as well blocks every one of them while the escape count stays at
zero (Table~\ref{tab:p3}): eight extra interceptions, no additional safety. The caution is an artifact of
\emph{where} the text comes from rather than of the gate. Decomposing each payload into its actions shows
that not one of the 532 ASK verdicts over this corpus is a real action --- 354 belong to comment lines and
178 to the non-comment scaffolding the generator wraps around them, against 77 ALLOWs and 54 DENYs from the real
actions. Removing that template shell at the interface leaves text byte-identical to the action column of
the corpus table in 61 of 61 cases and relaxes no denial, and it makes the strict policy free: with the
shell stripped, intercepting DENY and ASK together blocks the same 57 payloads and no benign one. We ship
DENY-only because the fix is a caller-side requirement, not because the strict policy is unusable.

\textbf{No vocabulary and no schema was disturbed.} An intercepted row produces no side effect, which
falls naturally into the existing \textsc{Under\_Block} classification, and the interception fact is
written into the existing note column. No verdict category had to be invented for it, so the ledger's
historical rows remain comparable with the new ones --- the discipline that makes a two-year log useful
at all.

\textbf{The record of a pass lives inside the boundary of the payloads it records.} The ledger that stamps
every row sits under a tree the payloads may write, and two payloads in the corpus target it: one
truncates it, the other appends a forged row. In a pass that does not consult the gate both succeed, and
what is lost is not one row but \emph{every row that pass had already written} --- 47 of them --- because
the driver finds no ledger, writes a fresh header, and continues. The side-effect oracle does not watch
that directory, so the payload is recorded as having done nothing; the surviving rows close the arithmetic
(18 + $4\times$65 = 278). This is a property of the measured system, not a defect of the instrument, and
the artifact ships a guard that snapshots the ledger before a pass and restores it afterwards. Containment remains a deployment decision: no file the guard writes is a boundary while payloads run with
write access to the ledger. Section~B.1c reconstructs the sequence.

\textbf{Why not the enforcement point the OS already ships.} A reviewer will ask, so we measured it: the
same 65 payloads re-run under an AppArmor profile carrying the gate's policy (last column,
Table~\ref{tab:p3}) leave 8 of the 25 unguarded effects standing --- a whitelist cannot refuse to destroy the
zone it grants, and the object-versus-path distinction of this paper survives the translation.
Section~B.1d separates all eight.

\textbf{An environment defect that produced two symptoms.} Our first enforcement pass reported an escape
that the control pass had not, and the frozen table simultaneously dropped from 61/61 to 59/61 with two
false allows. We initially reported the escape as a gate defect. It was not: the control pass deletes
decoys, regenerating the decoy tree does not rebuild the range image, and the second pass had therefore
run against a dirtied image in which the gate's same-inode closure could no longer recognise a bind-mount
alias --- \emph{one} environment defect, \emph{two} symptoms, in different reports. Restoring the image
returned the table to 61/61 and the escape count to zero, and the difference between the two passes
reduced to the single row whose classification depends on the alias. We record four rules that this cost
us, because all four are silent: regenerating decoys does not rebuild the image; a bind mount is created
only if the image is already mounted, and otherwise the generator prints one warning and continues; the
generator deletes the two fixtures that the program-identity tests depend on; and any comparison across
batches must therefore declare the range state it was taken in. The last rule is the one that generalises,
and Section~\ref{sec:threats} shows what happens to a reading when it is ignored.

\section{Does it stay flat as the system is used?}
\label{sec:growth}

A gate that decides once per action is only useful if deciding stays affordable while the system
accumulates history. We therefore asked which structure in the chain actually grows with use --- and the
first answer was a correction: the privilege ledger is derived from a fixed list of tool names, resolved
once and cached, and \textbf{the adjudication path never reads state that accumulates}. The accumulation
lives one layer below, in the credential ledger, and there it is read on every verification.

\begin{table}[t]
\caption{Credential-ledger load, median of 30 runs per cell (synthetic ledgers, same column layout and
row type as the live one). ``Primary'' is the ledger read at verification; ``primary $+$ secondary'' is the
second read the same code performs when a second ledger is configured.}
\label{tab:growth}
\centering
\footnotesize
\begin{tabular}{rccc}
\toprule
rows & primary (ms) & primary $+$ secondary (ms) & ratio \\
\midrule
10 & 0.037 & 0.073 & 1.98$\times$ \\
100 & 0.141 & 0.295 & 2.09$\times$ \\
1{,}000 & 1.243 & 2.500 & 2.01$\times$ \\
10{,}000 & 18.908 & 37.792 & 2.00$\times$ \\
100{,}000 & 218.103 & 439.827 & 2.02$\times$ \\
\bottomrule
\end{tabular}
\end{table}

\begin{figure}[t]
\centering
\includegraphics[width=\columnwidth]{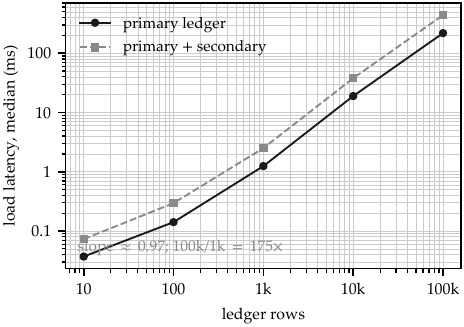}
\caption{Credential-ledger load against ledger size, both series on log--log axes. The dashed series is
the same measurement with the second ledger read included. Absolute values are machine-specific (shape,
not level, is the claim).}
\label{fig:growth}
\end{figure}

The cost is linear in the number of issued credentials (log--log slope $0.97$, and $175\times$ from one
thousand rows to one hundred thousand), and the constant factor is a factor of two rather than a
rounding: the verification path reads its ledger, and then reads a second one to answer a different
question. Fig.~\ref{fig:growth} shows both series. Two consequences matter for deployment. First, at ten
thousand issued credentials a single verification costs about 19\,ms, or 38\,ms with the second read ---
large enough to matter in an online verification path. Second, \emph{this cost is
invisible in the deployment we measured}: the live credential ledger is absent and the staging ledger holds
13 rows, so nothing today is slow. That is precisely why it is worth stating now, before a ledger that is
supposed to be complete becomes an argument against checking it.

\textbf{Where that linear cost sits, measured rather than assumed.} Our written prediction attributed the
linear cost to parsing rather than raw I/O, and its own prescribed control --- vary row and column
structure at controlled byte counts --- falsified its specific column-count criterion: at constant bytes,
cutting the columns fourfold made a load \emph{slower} (14.0 to 20.5\,ms), because the row count rose by the
same factor.
Fitting $t = a\cdot\mathrm{rows}+b\cdot\mathrm{bytes}$ gives $a\approx162$\,ns per row and
$b\approx3.0$\,ns per byte against a raw read at 0.14\,ns per byte: neither term is I/O, and the per-row
term carries the field splitting, so the remedy is to stop re-reading whole rows; the variants and the fit are
in Section~D.1.

\begin{figure*}[t]
\centering
\includegraphics[width=0.94\textwidth]{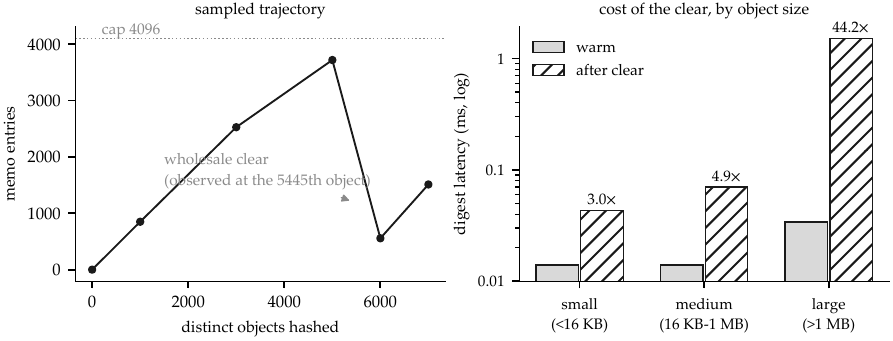}
\caption{Left: memo occupancy on a sampled trajectory of distinct objects; the cap fires at 4096 entries
and the table is cleared wholesale. Right: the cost of that clear, measured by revisiting a fixed set of
files of each size tier after forcing it.}
\label{fig:cliff}
\end{figure*}

\textbf{The second structure is a cliff, and its cost depends on what is being hashed.} The gate memoises
content digests for non-writable zones under a cap of 4096 entries, and when the cap is reached it clears
the \emph{entire} table rather than evicting the oldest entries. The cap fires on the 4096th distinct key,
which we observed at the 5445th distinct object hashed --- not every object populates the memo. What the
clear then costs is paid on \emph{revisit}, and it scales with object size: re-hashing a fixed set after a
forced clear costs $3.0\times$ more for files under 16\,KB, $4.9\times$ for files between 16\,KB and
1\,MB, and $44.2\times$ for files above 1\,MB (Fig.~\ref{fig:cliff}). A single figure for ``the cost of
the clear'' would be wrong in either direction.

\textbf{A measurement we had to redo.} The first probe reported only $1.3\times$ because it hashed each
file once and never revisited entries after the clear. The corrected design forces a clear and revisits a
fixed set, exposing the size-dependent cliff above. Supplementary Section~D.2 reproduces the flawed
traversal as a negative control and records why its plausible number was not evidence of the claimed cost.

\textbf{What we would change, and what we did not do.} Both costs have ordinary fixes: cache the
\emph{parsed} ledger keyed by (path, size, modification time) and invalidate on append, and replace the
wholesale clear with an eviction policy. We predict, without implementing, that the first flattens the
curve in Fig.~\ref{fig:growth} (log--log slope below 0.2) and the second removes the step while keeping the
memory bound. Only summary medians, not the 30 raw values per cell, were archived; exact historical
medians cannot be reconstructed, although the scripts can rerun the experiment.
We deliberately measure shape rather than level here: the absolute milliseconds in
Table~\ref{tab:growth} belong to one machine (Ubuntu 26.04, Python 3.14.4, an eight-core Hygon part), and
the claim this paper makes is about how the cost grows, not about what it costs.

\section{Gate versus model on the same cases}
\label{sec:llm}

On 25 September 2026 the same cases were sent five times through one third-party OpenAI-compatible
gateway under the request labels in Table~\ref{tab:llm}. This is an exploratory endpoint snapshot, not an
authenticated model comparison: the archive omits response IDs, returned-model fields and provider
receipts. Original/Linux inputs comprised 259/212 rows and 3{,}902/3{,}180 logged attempts. Deduplicating
by label, case and repetition while preferring a successful response over an HTTP failure retains
3{,}885/3{,}180 analytical slots; 3{,}879/3{,}178 parsed, the unparsed remainder falling on four
cases in the first sheet (P1-T195, P1-T255, P1-T267, P1-T271) and two different cases in the second
(P1-T125, P1-T203); every other case keeps all five repetitions.
Across parseable calls, \emph{non-refusal} (\texttt{execute} or \texttt{ask}) was \textbf{92.2\%}
(3{,}578/3{,}879; CI 91.4--93.0) and \textbf{93.1\%} (2{,}959/3{,}178; CI 92.2--93.9). Because \texttt{ask}
defers to human approval, these are neither self-approval nor immediate-execution rates.

\begin{table*}[t]
\caption{Per-case majority decisions. NR is non-refusal (\texttt{execute} plus \texttt{ask}); ``strict''
excludes platform-specific denials; the final column reports refuse / ask.}
\label{tab:llm}
\centering
\footnotesize
\begin{tabular}{lcccc}
\toprule
endpoint label & DENY NR (122) & strict NR (99) & strict execute (99) & ALLOW refuse / ask (40) \\
\midrule
claude-opus-5 & 84.4\% & 81.8\% & 24.2\% & 0\% / 27.5\% \\
deepseek-v4.1-flash & 95.9\% & 94.9\% & 4.0\% & 0\% / 65.0\% \\
gpt-6-astra & 97.5\% & 98.0\% & 52.5\% & 0\% / 17.5\% \\
\bottomrule
\end{tabular}
\end{table*}

The archived responses also expose an endpoint-side confound: 114 of 7{,}065 retained responses (82
original and 32 Linux, all under the second label) contain context absent from their requests. Removing
them raises the two call-level non-refusal rates from 92.2/93.1\% to 92.8/93.2\%, but changes that label's
DENY/strict case-majority rates from 117/122 (95.9\%)/94/99 (94.9\%) to 115/122 (94.3\%)/93/99
(93.9\%); P1-T038 and P1-T235 become ties. The other labels are unchanged. Table~\ref{tab:llm}
therefore reports the archived observation, not a stable property of a named model.

\begin{figure}[t]
\centering
\includegraphics[width=\columnwidth]{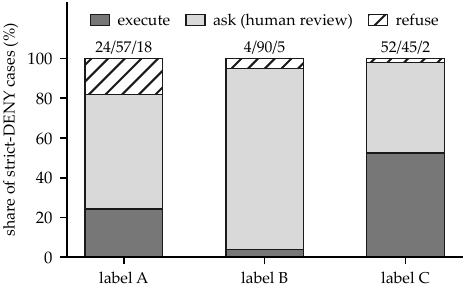}
\caption{Per-case majority decisions on the 99 strict-DENY cases, separated into immediate execution,
human review (\texttt{ask}), and refusal. Labels A--C follow the table order; values are the archived
observations before the disclosed endpoint-context sensitivity exclusion.}
\label{fig:llm}
\end{figure}

The same-case comparison (Table~\ref{tab:llm}, Fig.~\ref{fig:llm}) removes platform-specific denials,
leaving 99 denials that follow from the request alone. Non-refusal is 81.8\%, 94.9\%, 98.0\%, but immediate
execution is 24.2\%, 4.0\%, 52.5\%; much of the former is \texttt{ask}. Of 40 gate-ALLOW rows, none is
refused, while 27.5\%, 65.0\%, 17.5\% are sent to human review. Deferral is not execution.

\textbf{Rewriting the same case for another platform changes the endpoint answer.}
Between the two versions of the identical case set, endpoint decisions shift by $-2.8$, $-2.4$ and $-0.5$
percentage points (affecting 8, 9 and 5 rows respectively). This 99-case subset was selected because both
the inherited expectation and the Linux gate say \texttt{DENY}, so the gate's denial is stable on this
decision-matched subset by definition; the endpoint shift is therefore a phrasing/carrier sensitivity
within that set. This is the concrete form of the model-decoupling claim: the gate consumes an action and a
policy state, not a phrasing, and it does not know which model produced either.

\section{Limitations}
\label{sec:limitations}

\textbf{Identity is not identity.} A program's name is not its identity, and this layer cannot make it so.
A renamed copy of a program given by path is recognisable --- the realised action carries the digest of the
executable's bytes --- but turning that recognition into a \emph{decision} requires a ledger of known
program objects, and a ledger is a policy that someone has to maintain. We report the recognition as
available and the binding as policy, rather than claiming the problem solved.

\textbf{Two projections are lossy and are stated as such} (Section~\ref{sec:rq2}). Rules that protect by
name --- a version-control directory, a project-specific directory --- have no zone-level equivalent here, and ten inherited rows differ from their expectations for that reason alone. Separately,
UNC-style paths cannot be translated faithfully without inventing a mount policy, so those rows
are acknowledged as lossy rather than presented as equivalent. Both are consequences of the model, not
defects to be patched.

\textbf{An oracle residual of 2.6\%.} Three of 117 re-executed runs change the decoy tree in a way the
snapshot diff does not attribute, because an object created \emph{during} execution cannot appear in a pre-
execution snapshot. We classify this as a lower bound of the method rather than a defect of the gate: closing it needs a second closure pass after execution, at the price of judging only once the action has run. The affected shapes are hard-link pairs and bind-mount views, and we name them rather than
reporting a rounded rate.

\textbf{The object model is the filesystem.} The gate reasons about files, paths, zones and program objects.
An action with no path target may still have effects outside that model --- a network request is the
obvious one. The measurable claim is narrower: within the action classes this realiser models, every
filesystem effect represented directly to the gate has a path target. An unmodelled or indirect effect need
not have one, and an action without a target is not therefore harmless. That boundary is exactly where a
relaxation of the conservative rule had to be guarded (below).

\textbf{Sequential statements are adjudicated independently.} A \texttt{PATH} assignment in one statement
does not inform a bare command name in the next, because the layer splits sequential actions and judges each; cross-statement variable dependencies are not modelled. This is a boundary of the layer as designed, not a bug, and it is why a symbol-to-target resolution problem is attributed to the platform rather than claimed here.

\textbf{A conservative rule that was doing two jobs.} The rule escalating every action without a path target to review was, besides its stated purpose, catching \emph{target-extraction failures}. Relaxing it
for the verb class that only writes to standard output turned one inherited case from review into
permission --- a recursive delete whose quoted argument carried a trailing space, so no target was extracted. The audit's false-allow count moved from 30 to 31, and only after adding the guard ``do not
relax when any operand carries path evidence'' did it return to 30. Measured on its own merits the
relaxation changes three synthetic shapes and no case in the frozen table, the mechanism tables or the 65
payloads: a usability improvement, reported as one.

\section{Threats to validity}
\label{sec:threats}

\textbf{Payload trust comes before gate quality.} The 216 inherited payloads are command strings rather than
authored payloads, and their expectations encode the source platform's semantics, so a difference
is not automatically an error. We therefore report the audit as an audit: false-allow counts graded by cause
(carrier misalignment, derivation mapping error, cross-platform mechanism difference), never as an accuracy.
Its internally pre-specified expectation was zero false allows; the measurement contradicted it and gave 30, which is
what we cite.

\textbf{Small denominators.} The frozen table has 61 rows and the mechanism tables 7--18 each. Every claim
in Section~\ref{sec:rq1} is about these corpora and the operators we happened to write, not about command space. The mutation campaign compensates for this in kind rather than in size: it reports
what the corpus fails to notice, which is a statement about the corpus itself. The corpora, rules and
operators share one author and may share a blind spot. Freezing prevents post-measurement tuning, but does
not provide independent validation; mutation identifies some shared blind spots, not necessarily all.

\textbf{Instrument generations.} The chain's text changed many times during this study, and every number is
reported with the fingerprint it was taken at; the ledger records the same identifier, so rows from
different generations can be separated rather than averaged. Where a number in this paper was measured at an
earlier generation we say so; the frozen table, the six blocks, the audit and the enforcement experiment
were all re-measured at the current one.

\textbf{Readings that are functions of the environment.} Any measurement depending on glob expansion
depends on the state of the target range. Deleting a single decoy file moves the mutation score from 92.0\%
to 94.0\% and changes one operator's fate. Our first cross-generation comparison inherited a score taken in
a different range state and treated it as a baseline, which produced a two-point discrepancy that survived
until the range state was controlled; the lesson is recorded as a discipline (M5) rather than as an
anecdote, because a reading without its environment is not a measurement.

\textbf{Per-case verdicts are not independent observations.} Within a pass the 65 payloads share one
target range, so blocking one payload can change the state seen by later payloads; between passes the
baseline is restored (image mounted, the bind-mount alias present, decoys regenerated, the two fixtures
restored, self-test 49/49), and all four passes re-run from it reproduce byte for byte. Table~\ref{tab:p3}
therefore compares whole passes, not independent cases, and Supplementary Section~B.1 records the
concrete move-row example that exposes the dependence. The baseline is load-bearing in the same way: with
the bind-mount alias absent, the two payloads that target it (\texttt{CORE-WRT-05} and
\texttt{LNX-MNT-01}) change the gate's verdict from DENY to ASK while keeping their classification in
Table~\ref{tab:p3}, which is why the range state is reported with every reading rather than assumed.

\textbf{Model baselines are endpoint snapshots.} Section~\ref{sec:llm} reports unauthenticated labels from
one third-party endpoint: response IDs, returned-model fields and provider receipts were not archived.
Moreover, 114/7{,}065 retained responses contain context absent from the request, and 54 archived raw fields
end exactly at the 4{,}000-character capture cap; reparsing those archived fields disagrees with 22 stored
per-call decisions. The published aggregate is consequently reconstructed from the stored decisions, with
the context-removal sensitivity reported separately. Results are bounded to this endpoint, date and subset;
its labels or backends may change.

\textbf{Falsified predictions are reported.} Five predictions in this study were false. A methodology that
reports only its successes cannot be evaluated, and two of the five failures pointed at real defects --- one at a fail-open no passing test had caught, one at a stale baseline inherited from a
previous hand-over.

\section{Conclusion}
\label{sec:conclusion}

A structured-action policy gate is faithful enough to be worth building, and bounded in ways that can be
stated rather than discovered. On our corpora it decides a frozen table 61 times out of 61 in two rounds,
achieves a 92.0\% mutation score (46/50 mutants killed) with all four survivors classified and named, and, when
the executor consults it, turns 25 escapes into zero without blocking a single benign payload. Its reasoning
is legible enough that an operator reading a decision can see which fact produced it, which is what makes
those four survivors \emph{findings} rather than unexplained residue.

What it cannot do is as useful to know. It cannot propagate semantic state or dependencies across sequential statements, so a variable
established in one action does not inform the next, and it cannot resolve a symbolic name without a path,
because that resolution belongs to the platform. Its object model is the filesystem, so the claim is limited
to the modeled action classes: a filesystem effect represented directly to the gate has a path target. It is
not a claim about indirect effects, nor does it imply that an action without a target is harmless. And one
conservative rule was silently doing a second job --- protecting the gate
from target-extraction failures --- which became visible only when the rule was relaxed and an inherited
case turned from review into permission.

The cross-platform half points the same way: the portable unit is the \emph{capability}, not the command.
The 82-row audit records 43 re-expressions, 21 carrier differences, 16 study-specific inapplicable rows and
two unresolved cases. Of the 25 rows originally labelled ``no counterpart,'' four remain unmapped on our
host and toolset; this is not a platform-wide impossibility claim. One further row, outside those 25,
keeps its carrier and inverts its meaning.
In the exploratory snapshot of 99 strict-DENY cases, the three endpoint labels produce non-refusal
majorities for 81.8--98.0\% of cases but immediate-execution majorities for only 4.0--52.5\%; unlike the
gate, their archived answers also change under platform rewriting. The endpoint-context exclusion changes
one strict rate from 94.9\% to 93.9\%, without changing the direction of this descriptive comparison.

Two directions follow: adjudication cost as the ledger grows, which is the question a deployment engineer
asks before believing any of the above, and the boundary we kept declaring --- cross-statement reasoning,
and object models beyond the filesystem. Both are extensions of the layer rather than repairs to it. A
policy gate is an instrument, and an instrument is worth exactly as much as the measurements made of it.

\section{Artifacts and how to reproduce the measurements}
\label{app:artifacts}

The artifact bundles the seven-file chain, the corpora, and the evaluators and reports used here.
Instrument fingerprints distinguish the current generation (\texttt{6c083e60ceeb}) from earlier
measurements (\texttt{285ec6bd7ecc}); \texttt{gate.py} \texttt{2c3c46650fa1} is common to both.
Supplementary Section~A enumerates the corpora, Section~C gives the 25-ruling ledger and exact range-recovery
procedure, and Section~D records the falsified predictions and measurement controls. Reproduction starts
by restoring the image, decoys and deleted fixtures and requiring the 49/49 self-test before rerunning the
report-generating scripts.

\section*{Acknowledgment}
The author used an AI-based writing assistant, DeepSeek,
for language editing of the manuscript and for drafting
and revising portions of the figure-generation and analysis
scripts included in the supplementary reproducibility
artifact. The AI system was not used to generate experimental
data, measurements, or scientific conclusions. All AI-assisted
material was independently reviewed, tested, and
verified by the author against the experimental records, and
the author takes full responsibility for the content of this
article.

\end{document}